\pdfoutput=1
\documentclass[twocolumn]{aastex62}

\received{xxx}
\revised{yyy}
\accepted{zzz}

\submitjournal{AAS Journals}

\shorttitle{LXUV History of TRAPPIST-1}
\shortauthors{Fleming et al.}

\usepackage{hyperref}
\usepackage{xspace}
\usepackage{graphicx}
\usepackage{amsmath}
\usepackage[caption=false]{subfig}
\usepackage[ruled,vlined]{algorithm2e}

\def\gsim{~\rlap{$>$}{\lower 1.0ex\hbox{$\sim$}}}
\def\lsim{~\rlap{$<$}{\lower 1.0ex\hbox{$\sim$}}}

\newcommand{\vplanet}[0]{\texttt{VPLanet}\xspace}
\newcommand{\emcee}[0]{\texttt{emcee}\xspace}
\newcommand{\approxposterior}[0]{\texttt{approxposterior}\xspace}

\newcommand{\stellar}[0]{\texttt{STELLAR}\xspace}

\begin{document}

\title{On The XUV Luminosity Evolution of TRAPPIST-1}



\author[0000-0001-9293-4043]{David P. Fleming}
\affil{Astronomy Department, University of Washington \\
Box 951580, Seattle, WA 98195}
\affil{NASA NExSS - Virtual Planetary Laboratory Lead Team, USA}

\author{Rory Barnes}
\affiliation{Astronomy Department, University of Washington \\
Box 951580, Seattle, WA 98195}
\affil{NASA NExSS - Virtual Planetary Laboratory Lead Team, USA}

\author[0000-0002-0296-3826]{Rodrigo Luger}
\affil{NASA NExSS - Virtual Planetary Laboratory Lead Team, USA}
\affiliation{Center for Computational Astrophysics, Flatiron Institute \\
New York, NY 10010}

\author[0000-0002-9623-3401]{Jacob T. VanderPlas}
\affiliation{Google \\
601 N 34th St, Seattle, WA 98103}


\begin{abstract}

We model the long-term XUV luminosity of TRAPPIST-1 to constrain the evolving high-energy radiation environment experienced by its planetary system. Using Markov Chain Monte Carlo (MCMC), we derive probabilistic constraints for TRAPPIST-1's stellar and XUV evolution that account for observational uncertainties, degeneracies between model parameters, and empirical data of low-mass stars. We constrain TRAPPIST-1's mass to $m_{\star} = 0.089 \pm{0.001}$ M$_{\odot}$ and find that its early XUV luminosity likely saturated at $\log_{10}(L_{XUV}/L_{bol}) = -3.03^{+0.23}_{-0.12}$. From the posterior distribution, we infer that there is a ${\sim}40\%$ chance that TRAPPIST-1 is still in the saturated phase today, suggesting that TRAPPIST-1 has maintained high activity and $L_{XUV}/L_{bol} \approx 10^{-3}$ for several Gyrs. TRAPPIST-1's planetary system therefore likely experienced a persistent and extreme XUV flux environment, potentially driving significant atmospheric erosion and volatile loss. The inner planets likely received XUV fluxes ${\sim}10^3 - 10^4\times$ that of the modern Earth during TRAPPIST-1's ${\sim}1$ Gyr-long pre-main sequence phase. Deriving these constraints via MCMC is computationally non-trivial, so scaling our methods to constrain the XUV evolution of a larger number of M dwarfs that harbor terrestrial exoplanets would incur significant computational expenses. We demonstrate that \approxposterior, an open source Python machine learning package for approximate Bayesian inference using Gaussian processes, accurately and efficiently replicates our analysis using $980\times$ less computational time and $1330\times$ fewer simulations than MCMC sampling using \emcee. We find that \approxposterior derives constraints with mean errors on the best fit values and $1\sigma$ uncertainties of $0.61\%$ and $5.5\%$, respectively, relative to \emcee.

\end{abstract}




\section{Introduction} \label{sec:intro}

The James Webb Space Telescope (JWST) is poised to detect and characterize the first terrestrial exoplanet atmospheres via transmission spectroscopy. This search will likely focus on planets orbiting nearby M dwarfs given their favorable relative transit depths, the potential buildup of biosignature gases due to UV-driven photochemisty \citep{Segura2005}, and the large occurrence rates of M dwarf planets \citep{Dressing2015}. The correct interpretation of those observations, however, is predicated on understanding the system's long-term evolution, most importantly processes that could impact the planet's atmospheric state and habitability, such as atmospheric escape, water loss, and the potential buildup of an abiotic O$_2$ atmosphere \citep{Watson1981,Lammer2003,MurrayClay2009,Luger2015}. These volatile escape mechanisms are partially driven by the host star's XUV luminosity (X-ray and EUV emission ranging over approximately 1-1000\AA), and therefore characterizing the long-term stellar XUV evolution of late M-dwarfs is critical to assessing the present state of their planets, including habitability.

High-energy stellar radiation originates from the corona via the heating of magnetically-confined plasma \citep{Vaiana1981}. The stellar magnetic field is likely generated via differential rotation within the stellar convective envelope \citep{Parker1955}, linking rotation to stellar activity and XUV emission. Stellar rotation rates slow over time due to magnetic braking \citep{Skumanich1972}, causing XUV emission to decline with time. The X-ray luminosity ($L_{X}$) of FGK stars, for example, has been empirically shown to monotonically decrease with age \citep{Jackson2012}. This trend has also been observed for commonly-used proxies for stellar age, rotation period and Rossby number \citep[Ro = $P_{rot}/\tau$ for convective turnover timescale $\tau$,][]{Pizzolato2003,Wright2011}. 

Stellar activity evolution is characterized by two distinct phases. First, in the saturated phase, young, rapidly-rotating stars ($\mathrm{Ro}\lsim 0.1$) maintain a constant $L_{X}/L_{bol} \approx 10^{-3}$ \citep{Wright2011,Jackson2012}. Then, at longer rotation periods and larger Ro, stars transition to the unsaturated phase in which $L_{X}/L_{bol}$ exponentially decays over time \citep{Pizzolato2003,Ribas2005}. Recent work has shown that the stellar dynamo processes that generate magnetic fields and drive XUV emission in fully-convective M dwarfs follow the same evolution with Ro as described above for solar-type stars \citep{Wright2016,Wright2018}. We can therefore apply this model to examine the XUV evolution of individual fully-convective stars.

TRAPPIST-1 \citep{Gillon2016,Gillon2017}, an ultracool dwarf located 12 pc from Earth, harbors 7 approximately Earth-sized transiting planets that are prime targets for JWST transmission spectroscopy observations \citep{Morley2017,Lincowski2018,Lustig2019}. TRAPPIST-1's high observed L$_{X}$ \citep{Wheatley2017}, short photometric rotation period \citep[3.3 d, ][]{Luger2017}, and low Rossby number \citep[Ro $\approx 0.01$, ][]{Roettenbacher2017} suggest that TRAPPIST-1 is still saturated today \citep{Pizzolato2003,Wright2011,Wright2018}. Both \citet{Roettenbacher2017} and \citet{Morris2018} suggest that the photometrically-determined rotation period is inaccurate, with the latter study proposing that the 3.3 d period corresponds to a characteristic timescale for active regions on the stellar surface. TRAPPIST-1's $v \sin i = 6$ km s$^{-1}$ \citep{Barnes2014}, however, implies a rotation period of ${\sim}1$ d for $i = 90^{\circ}$, providing evidence that TRAPPIST-1's rapid rotation is physical and consistent with saturation \citep[$P_{rot} \lsim 20$ d,][]{Wright2018}. 

The TRAPPIST-1 planetary system currently receives significant high-energy fluxes \citep{Bourrier2017b,Wheatley2017,Peacock2019}, possibly a consequence of TRAPPIST-1 remaining in the saturated regime. These fluxes were likely more extreme during the pre-main sequence, driving significant water loss and potentially rendering the planets uninhabitable \citep{Bolmont2017,Bourrier2017a}. Here, we model the long-term stellar and XUV evolution of TRAPPIST-1 to characterize the evolving XUV environment of its planetary system. We use MCMC to derive probability distributions for our model parameters that describe the XUV evolution that are consistent with TRAPPIST-1's observed properties and their uncertainties.

TRAPPIST-1 is not the only system that merits this modelling, however, as the Transiting Exoplanet Survey Satellite will likely discover additional transiting planets orbiting in the habitable zone of nearby M dwarfs \citep{Barclay2018}, some of which may be suitable targets for atmospheric characterization with JWST. In this work, we show that stellar XUV histories can be accurately inferred using machine learning \citep[\approxposterior, ][]{FlemingVanderPlas2018}, but using $980\times$ less computational resources than traditional MCMC methods. This speed-up enables our methods to scale to additional stars that host potential targets for atmospheric characterization and is generalizable to a large number of applications, potentially enabling Bayesian statistical analyses that are otherwise intractable with traditional MCMC approaches, e.g. \emcee \citep{ForemanMackey2013}.

We describe our model and statistical methods in $\S$~\ref{sec:methods}. We present our results and demonstrate the ability of machine learning to reproduce our analysis in $\S$~\ref{sec:results}, and discuss the implications of our results in $\S$~\ref{sec:discussion}. In $\S$~\ref{sec:app}, we describe the \approxposterior algorithm and discuss its convergence properties.

\section{Methods} \label{sec:methods}

\subsection{Simulating XUV Evolution with \vplanet} \label{sec:model}

We simulate TRAPPIST-1's stellar evolution using the \stellar module in \vplanet\footnote{\vplanet is publicly available at \href{https://github.com/VirtualPlanetaryLaboratory/vplanet}{https://github.com/VirtualPlanetaryLaboratory/vplanet}.} \citep{Barnes2019}, which performs a bicubic interpolation over mass and age of the \citet{Baraffe2015} stellar evolution tracks. The \citet{Baraffe2015} models (also employed by both \citet{Burgasser2017} and \citet{vanGrootel2018} to constrain TRAPPIST-1's stellar properties) were computed for solar metallicity stars and hence are suitable for TRAPPIST-1 whose [Fe/H] is consistent with solar \citep[][see also \citet{Burgasser2017}]{Gillon2016}.

We assume TRAPPIST-1's L$_{XUV}$ evolution traces that of L$_{X}$ and use the \citet{Ribas2005} model,
\begin{align}
\label{eqn:lxuv}
\frac{L_\mathrm{XUV}}{L_\mathrm{bol}} = \left\{
				\begin{array}{lcr}
					f_\mathrm{sat} &\ & t \leq t_\mathrm{sat} \\
					f_\mathrm{sat}\left(\frac{t}{t_\mathrm{sat}}\right)^{-\beta_\mathrm{XUV}} &\ & t > t_\mathrm{sat}
				\end{array}
				\right.,
\end{align}
where $f_{sat}$ is the constant ratio of stellar XUV to bolometric luminosity during the saturated phase, $t_{sat}$ is the duration of the saturated phase, and $\beta_{XUV}$ is the exponent that controls how steeply L$_{XUV}$ decays after saturation. In practice, we define $f_{sat} = \log_{10}(L_{XUV}/L_{bol})$ and transform Eqn.~(\ref{eqn:lxuv}) accordingly.

Note that each \vplanet simulation (and hence likelihood calculation, see $\S$~\ref{sec:mcmc:like}) in principle only requires interpolating the \citet{Baraffe2015} $L_{bol}$ tracks and evaluating an explicit function of time to compute $L_{XUV}$, both computationally-cheap tasks. \vplanet, however, is a general purpose code designed to simulate the evolution of an exoplanetary system and its host star by simultaneously integrating coupled ordinary differential equations and explicit functions of time that describe the evolution. This generalized structure requires numerous steps to facilitate physical couplings, such as validation steps and a host of intermediate calculations \citep[for more details, see][]{Barnes2019}. Moreover, \stellar simultaneously evolves a star's radius, effective temperature, radius of gyration, $L_{XUV}$, and rotation rate in addition to $L_{bol}$, adding computational overhead. Each \vplanet simulation using \stellar therefore lasts about 10s.


\subsection{Markov Chain Monte Carlo Analysis} \label{sec:mcmc}

We use \texttt{emcee}, a Python implementation of the affine-invariant Metropolis-Hastings MCMC sampling algorithm \citep{ForemanMackey2013}, to infer posterior probability distributions for our model parameters that describe the evolution of TRAPPIST-1. These distributions are conditioned on observations of TRAPPIST-1, the activity evolution of late-type stars, and account for both observational uncertainties and correlations between parameters. Our model parameters that we fit for via MCMC comprise the state vector
\begin{equation} \label{eqn:state}
    \textbf{x} = \{m_{\star}, f_{sat}, t_{sat}, \mathrm{age}, \beta_{XUV}\},
\end{equation}
where $m_{\star}$ and age are the stellar mass and age, respectively, and the other parameters are defined by Eqn.~(\ref{eqn:lxuv}). All of the code used to perform the simulations and analysis in this work is publicly available online.\footnote{ \href{https://github.com/dflemin3/trappist}{https://github.com/dflemin3/trappist}}

\subsection{Prior Probability Distributions} \label{sec:mcmc:priors}

Since we have few available observable properties of TRAPPIST-1 to use to condition our analysis ($L_{bol}$ and $L_{XUV}/L_{bol}$, see $\S$~\ref{sec:mcmc:like}), our prior probability distributions will strongly impact our results. We use previous studies and empirical data of late M dwarfs to assemble the best available constraints to serve as priors for our MCMC analysis. We list our adopted prior probability distributions in Table~\ref{tab:priors}.

Following \citet{vanGrootel2018}, we rely on TRAPPIST-1's luminosity and age to constrain its mass. We therefore adopt a simple uniform prior of $m_{\star} \sim \mathcal{U}(0.07, 0.11)$ M$_{\odot}$. For the age, we use the empirical estimate for TRAPPIST-1 derived by \citet{Burgasser2017}, age $\sim \mathcal{N}(7.6, 2.2^2)$ Gyr, as their thorough analysis considered both observations of TRAPPIST-1 and a host of empirical age indicators for ultracool dwarfs. This age distribution is consistent with \citet{Gonzales2019} who conclude that TRAPPIST-1 is a field-age dwarf based on their spectral energy distribution modeling. We cap the maximum age we consider at 12 Gyr. Younger ages have been suggested based on TRAPPIST-1's activity \citep[e.g.~$\gsim 500$ Myr,][]{Bourrier2017b}, but here we argue that behavior is consistent with an extended saturation timescale.

We construct an empirical $f_{sat} = \log_{10}(L_{XUV}/L_{bol})$ distribution from the sample of fully-convective, saturated M dwarfs with observed $L_{X}$ from \citet{Wright2011}. For each star in the \citet{Wright2011} sample, we follow \citet{Wheatley2017} and estimate $L_{XUV}$ as a function of L$_{X}$ using Eqn.~(2) from \citet{Chadney2015}. We find that the distribution is well-approximated by a normal distribution, $f_{sat} \sim \mathcal{N}(-2.92, 0.26^2)$, and we adopt it as our prior.  

The duration of the saturated phase is estimated to be $t_{sat} \approx 100$ Myr for FGK stars \citep{Jackson2012}. Studies of stellar activity of late type stars as a function of stellar age, or its proxy, rotation period, indicate that the activity lifetime, and hence duration of the saturated phase, is likely longer for later-type stars \citep{Shkolnik2014,Wright2011,West2015}, with fully-convective M dwarfs potentially remaining active throughout their lifetimes \citep[$t_{sat} \gsim 7$ Gyr,][]{West2008,Schneider2018}. Furthermore, the spin-down timescales of late M dwarfs increases with decreasing stellar mass \citep{Delfosse1998}, with late M dwarfs retaining rapid rotation longer than earlier-type stars and hence remaining active for up to $P_{rot} \approx 86$ d \citep{West2015}, much longer than TRAPPIST-1's estimated rotation period. Given these constraints, we adopt a broad uniform $t_{sat}$ prior distribution capped by the maximum age we consider, $t_{sat} \sim \mathcal{U}(0.1, 12)$ Gyr. 

In the unsaturated phase, $L_{X}$, and hence $L_{XUV}$, decay exponentially with power law slope $\beta_{XUV}$ \citep{Ribas2005}. \citet{Jackson2012} find that $\beta_{XUV}$ does not significantly vary with stellar mass in their sample of FGK stars. Since \citet{Wright2016} found that the X-ray evolution of fully-convective stars is qualitatively similar to that of partially-convective FGK stars, we adopt the $\beta_{XUV}$ distribution of late K dwarfs from the \citet{Jackson2012} sample as our prior, $\beta_{XUV} \sim \mathcal{N}(-1.18, 0.31^2)$.

\begin{deluxetable}{lcc}
\tabletypesize{\small}
\tablecaption{Prior Distributions \label{tab:priors}}
\tablewidth{0pt}
\tablehead{
\colhead{Parameter [units]} & \colhead{Prior} & \colhead{Notes}
}
\startdata
$m_\star$ [$M_{\odot}$] & $\mathcal{U}(0.07, 0.11)$ & -- \\  
$f_{sat}$ & $\mathcal{N}(-2.92, 0.26^2)$ & \citet{Wright2011}  \\
$t_{sat}$ [Gyr] & $\mathcal{U}(0.1, 12)$ & -- \\
age [Gyr] & $\mathcal{N}(7.6, 2.2^2)$ & \citet{Burgasser2017} \\
$\beta_{XUV}$ & $\mathcal{N}(-1.18, 0.31^2)$ & \citet{Jackson2012}
\enddata 
\end{deluxetable}

\subsection{Likelihood Function and Convergence} \label{sec:mcmc:like}

We further condition our analysis on TRAPPIST-1's observed bolometric luminosity, $L_{bol} = 5.22 \pm{0.19} \times 10^{-4} \ L_{\odot}$ \citep[][but see also \citet{Gonzales2019}]{vanGrootel2018}, and $L_{XUV}/L_{bol} = 7.5 \pm{1.5} \times 10^{-4}$ \citep{Wheatley2017}. In other words, we require that our forward models (\vplanet simulations) yield results that are consistent with the observations of TRAPPIST-1 and their uncertainties. 

For a given state vector \textbf{x}, we define the natural logarithm of our likelihood function, $\ln \mathcal{L}$, as
\small
\begin{equation} \label{eqn:lnlike}
    \ln \mathcal{L} \propto -\frac{1}{2} \left[ \frac{(L_{bol} - L_{bol}(\textbf{x}))^2}{\sigma_{L_{bol}}^2} + \frac{(L_{XUV}/L_{bol} - L_{XUV}/L_{bol}(\textbf{x}))^2}{\sigma_{L_{XUV}/L_{bol}}^2} \right] \\
\end{equation}
\normalsize
where $L_{bol}$, $L_{XUV}/L_{bol}$ and $L_{bol}(\textbf{x})$, $L_{XUV}/L_{bol}(\textbf{x})$ are the observed values and \vplanet outputs given \textbf{x}, respectively, and $\sigma_{L_{bol}}$ and $\sigma_{L_{XUV}/L_{bol}}$ are the observational uncertainties. For each \textbf{x}, we compute the natural logarithm of the posterior probability at $\textbf{x}$, lnprobability, required for ensemble MCMC sampling as $f(\textbf{x}) = \ln \mathcal{L}(\textbf{x}) + \ln \mathrm{Prior}(\textbf{x})$. We use the distributions described in $\S$~\ref{sec:mcmc:priors} to calculate the natural logarithm of the prior probability of \textbf{x}, $\ln \mathrm{Prior}(\textbf{x})$. 

We run our MCMC with 100 parallel chains for 10,000 iterations, initializing each chain by randomly sampling each element of \textbf{x} from their respective prior distributions. During each step of the MCMC chain, \vplanet takes \textbf{x} as input and simulates TRAPPIST-1's evolution up to the age in \textbf{x}, predicting $L_{bol}$ and $L_{XUV}/L_{bol}$ to evaluate $\ln \mathcal{L}$. We discard the first 500 iterations as burn-in and assess the convergence of our MCMC chains by computing the integrated autocorrelation length and acceptance fraction for each chain. We find a mean acceptance fraction of 0.48 and a minimum and mean number of iterations per integrated autocorrelation length of 93 and 132, respectively, indicating that our chains have converged \citep{ForemanMackey2013}. Given our integrated autocorrelation lengths, our MCMC chain yielded about 10,000 effective samples from the posterior distribution.

\subsection{Inference with \approxposterior} \label{sec:methods:approx}

The methods presented above can be applied to any late-type star to constrain its $L_{XUV}$ history, given suitable priors and observational constraints. Our MCMC analysis, however, required 4,070 core hours on the University of Washington's Hyak supercomputer to converge. The main computational cost is incurred by running a ${\sim}10$s \vplanet simulation each MCMC step to evaluate $\ln \mathcal{L}$, requiring ${\sim}1,000,000$ simulations in total for the full MCMC analysis. Assuming similar convergence properties, repeating this analysis for even a modest sample of 30 stars would require~${\sim} 122,000$ core-hours, a significant computational expense. Moreover, performing a similar analysis with a more computationally-expensive model would only exacerbate this issue.

To mitigate the computational cost, we apply \approxposterior\footnote{\approxposterior is publicly available at \href{https://github.com/dflemin3/approxposterior}{https://github.com/dflemin3/approxposterior}.}, an open source Python machine learning package \citep{FlemingVanderPlas2018}, to compute an accurate approximation to the true MCMC-derived posterior distribution for TRAPPIST-1's XUV evolution. \approxposterior, a modified implementation of the ``Bayesian Active Learning for Posterior Estimation" (BAPE) algorithm of \citet{Kandasamy2017}, trains a Gaussian process (GP, see \citet{Rasmussen2006}) replacement for the lnprobability evaluation, learning on the results of \vplanet simulations. The GP is then used within an MCMC sampling algorithm, e.g. \emcee, to quickly obtain the posterior distribution. In our case, predicting the lnprobability using the GP (${\sim} 130 \mu$s) is $80,000 \times$ faster than running \vplanet (10s) each lnprobability evaluation, yielding a massive reduction in computational cost.

Following \citet{Kandasamy2017}, \approxposterior iteratively improves the GP's predictive ability by identifying high-likelihood regions in parameter space, and hence high posterior density regions, where the GP predictions are uncertain. \approxposterior then evaluates \vplanet in those regions to supplement the training set, improving the GP's predictive ability in the relevant regions of parameter space, while minimizing the number of forward model evaluations required for suitable predictive accuracy. Similar techniques using a GP surrogate model have been shown to rapidly and accurately infer Bayesian posterior distributions for computationally-expensive cosmology studies \citep[e.g.][]{Bird2019,McClintock2019}. 

To model the covariance between points in the GP training set, we use a squared exponential kernel,
\begin{equation} \label{eqn:kernel}
k(x_i, x_j) = \exp \left( - \frac{(x_i - x_j)^2}{2l^2} \right),
\end{equation}
where $x_i$ and $x_j$ are two arbitrary points in parameter space and $l$ is a hyperparameter that controls the scale length of the correlations. We assume correlations in each dimension have different scale lengths and fit for each $l$ by optimizing the GP's marginal likelihood of the training set data using Powell's method \citep{Powell1964}, randomly restarting this optimization 10 times to mitigate the influence of local extrema. To ensure our solution is numerically stable, we add a small white noise term of $\ln(\sigma_{\mathrm{w}}) = -15$ to the diagonal of the GP covariance matrix. 

We initially trained the GP on a set of 50 \vplanet simulations with initial conditions sampled from our prior distributions. We then ran \approxposterior until it converged after 7 iterations. Each iteration, \approxposterior selected 100 new training points according to the \citet{Kandasamy2017} point selection criterion. \approxposterior ran \vplanet at each point for a total of 750 training samples. The trained GP was then used within \emcee to quickly obtain the approximate posterior distribution following the same MCMC sampling procedure described above. In $\S$~\ref{sec:app}, we provide additional information about the \approxposterior algorithm and its convergence properties.


\section{Results} \label{sec:results}

\subsection{The Evolution of TRAPPIST-1}

\begin{figure*}[t]
\centering
	\includegraphics[width=0.75\textwidth]{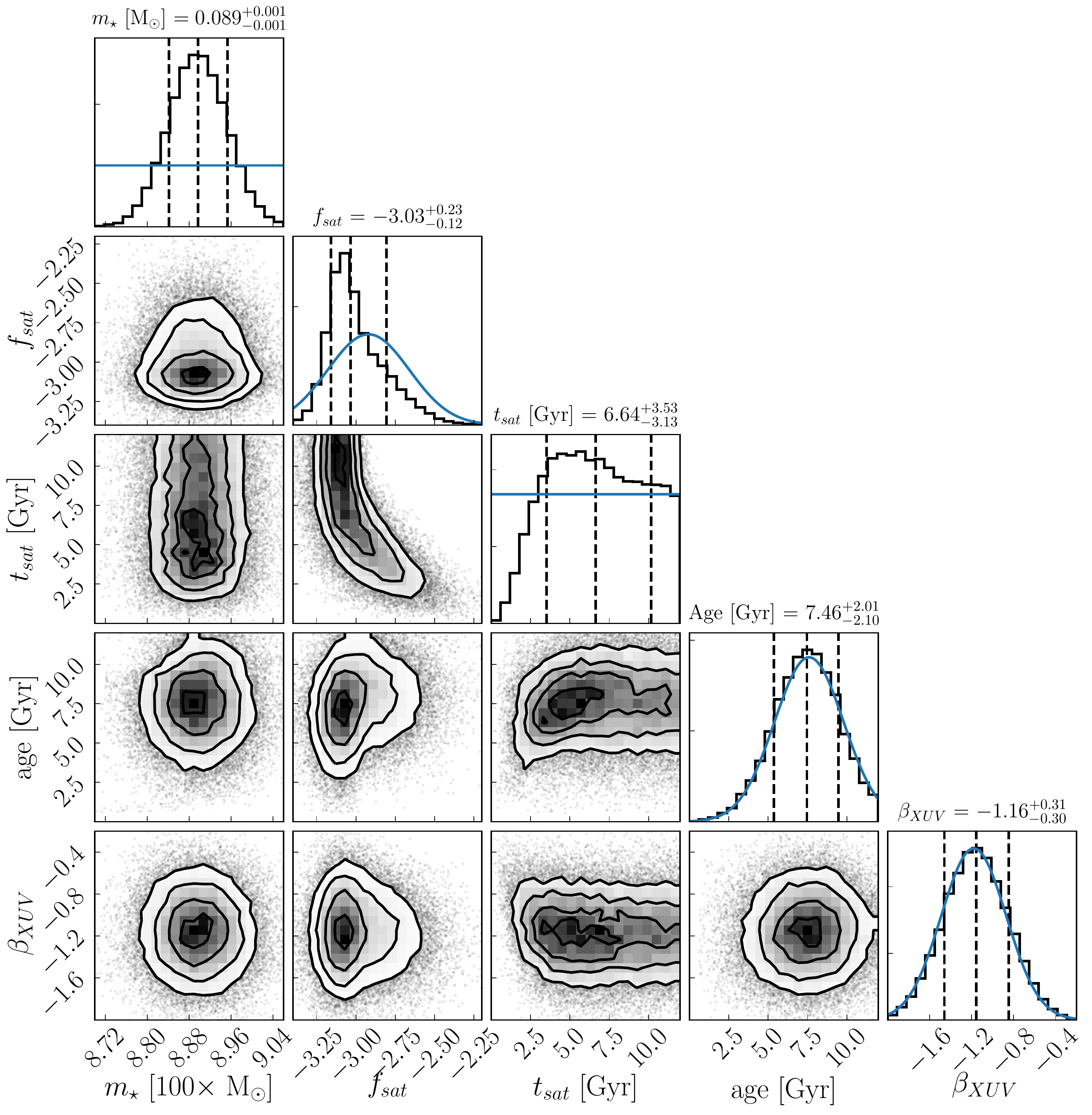}
   \caption{Joint and marginal posterior probability distributions for the TRAPPIST-1 stellar parameters given in Eqn.~(\ref{eqn:state}) made using \texttt{corner} \citep{ForemanMackey2016}. The black vertical dashed lines on the marginal distributions indicate the median values and lower and upper uncertainties from the 16th and 84th percentiles, respectively. The blue curves superimposed on the marginal distributions display the adopted prior probability distribution for each parameter. From the posterior, we infer that there is a $40\%$ chance that TRAPPIST-1 is still in the saturated phase today.}%
    \label{fig:corner}%
\end{figure*}

In Fig.~\ref{fig:corner}, we display the posterior probability distributions for our model parameters derived using MCMC with \vplanet and \emcee. We adopt the median values of the marginal distributions as our best-fit solutions and derive the lower and upper uncertainties using the 16th and 84th percentiles, respectively. We list these values in Table~\ref{tab:constraints}.

TRAPPIST-1 likely maintained a large $L_{XUV}$ throughout its lifetime as we find $f_{sat} = -3.03^{+0.23}_{-0.12}$ and $t_{sat} = 6.64^{+3.53}_{-3.13}$ Gyr, consistent with observed $L_{XUV}/L_{bol}$ and long activity lifetimes of late M dwarfs \citep{West2008,Wright2018}. The long upper-tail in the marginal $f_{sat}$ distribution arises from the combination of the degeneracy between $f_{sat}$ and $t_{sat}$ and from our strong empirical $f_{sat}$ prior that disfavors $f_{sat} \gsim -2.5$. The degeneracy stems from our model attempting to match TRAPPIST-1's observed $L_{XUV}/L_{bol}$. For example, larger values of $f_{sat}$ produce high initial $L_{XUV}/L_{bol}$, requiring shorter $t_{sat}$, and hence an earlier transition to unsaturated $L_{XUV}/L_{bol}$ decay, to decrease $L_{XUV}/L_{bol}$ to its observed value, and vice versa. 

Although our $t_{sat}$ prior distribution, based on empirical measurements of late M-dwarfs (see $\S$~\ref{sec:mcmc:priors}), equally favors both short and long saturation timescales, the marginal posterior density for $t_{sat}$ steeply declines for $t_{sat} \lsim 4$ Gyr. This decline implies that ultracool dwarfs like TRAPPIST-1 likely remain saturated for many Gyrs. Our analysis strongly disfavors short saturation timescales, with only a $0.5\%$ chance that $t_{sat} \leq 1$ Gyr, the saturation timescale adopted by \citet{Luger2015} in their analysis of water loss from exoplanets orbiting in the habitable zone of late M dwarfs and in \citet{Lincowski2018}. From the posterior distribution, we infer that there is a $40\%$ chance that TRAPPIST-1 is still in the high-$L_{XUV}/L_{bol}$ saturated phase today, suggesting that the TRAPPIST-1 planets could have undergone prolonged volatile loss.

The marginal age and $\beta_{XUV}$ posterior distributions reflect their prior distributions as for the former, $L_{bol}$ is not sufficient to constrain TRAPPIST-1's age beyond our adopted prior because the luminosities of ultracool dwarfs do not significantly change during the main sequence \citep{Baraffe2015}. The marginal posterior for $\beta_{XUV}$ does not vary from the prior because our XUV model is over-parameterized with 3 parameters to fit 2 observations, although all are motivated by empirical data and hence merit inclusion. Our model prefers to exploit the degeneracy between $f_{sat}$ and $t_{sat}$ to match TRAPPIST-1's observed $L_{XUV}$ in our MCMC instead of varying the slope of the unsaturated $L_{XUV}$ decay. Even though our model is over-parameterized, the observations of TRAPPIST-1 used to condition our probabilistic model do in fact influence the posterior distribution as the reduction in posterior variance relative to the prior can be seen in the joint posterior and marginal distributions of Fig.~\ref{fig:corner} for $m_{\star}$, $f_{sat}$, and $t_{sat}$.

In the joint posterior distribution, age and $\beta_{XUV}$ weakly correlate with $f_{sat}$, requiring a narrow spread of $f_{sat} \approx -3.05$ for young ages and steeper $\beta_{XUV}$, respectively. $\beta_{XUV}$ and $t_{sat}$ are uncorrelated, except at short $t_{sat}$ where steep $\beta_{XUV}$ are disfavored as this evolution would underpredict the observed $L_{XUV}$. We constrain TRAPPIST-1's mass to $m_{\star} = 0.089 \pm{0.001}$ M$_{\odot}$, in a good agreement with and $6\times$ more precise than the value derived by \citet{vanGrootel2018}. In $\S$~\ref{sec:evol}, we consider how this mass constrain impacts TRAPPIST-1's predicted radius.

Finally, we estimate the Monte Carlo standard error (MCSE) for each model parameter. The MCSE does not reflect the inherent probabilistic uncertainty in our model that arises from conditioning on data with uncertainties, but rather it approximates the error incurred by estimating parameters using an ensemble of MCMC chains of finite length. Using the batch means method \citep{Flegal2008,Flegal2010}, we find MCSEs for $m_{\star}$, $f_{sat}$, $t_{sat}$, age, and $\beta_{XUV}$ of $3.41 \times 10^{-6}$, $1.23 \times 10^{-3}$, $2.0 \times 10^{-2}$, $1.30 \times 10^{-2}$, and $2.12 \times 10^{-3}$, respectively. These errors are much less than the posterior uncertainty and can be safely ignored.


\subsection{Comparison with \approxposterior} \label{sec:approx}


\begin{figure*}
\centering
	\includegraphics[width=0.75\textwidth]{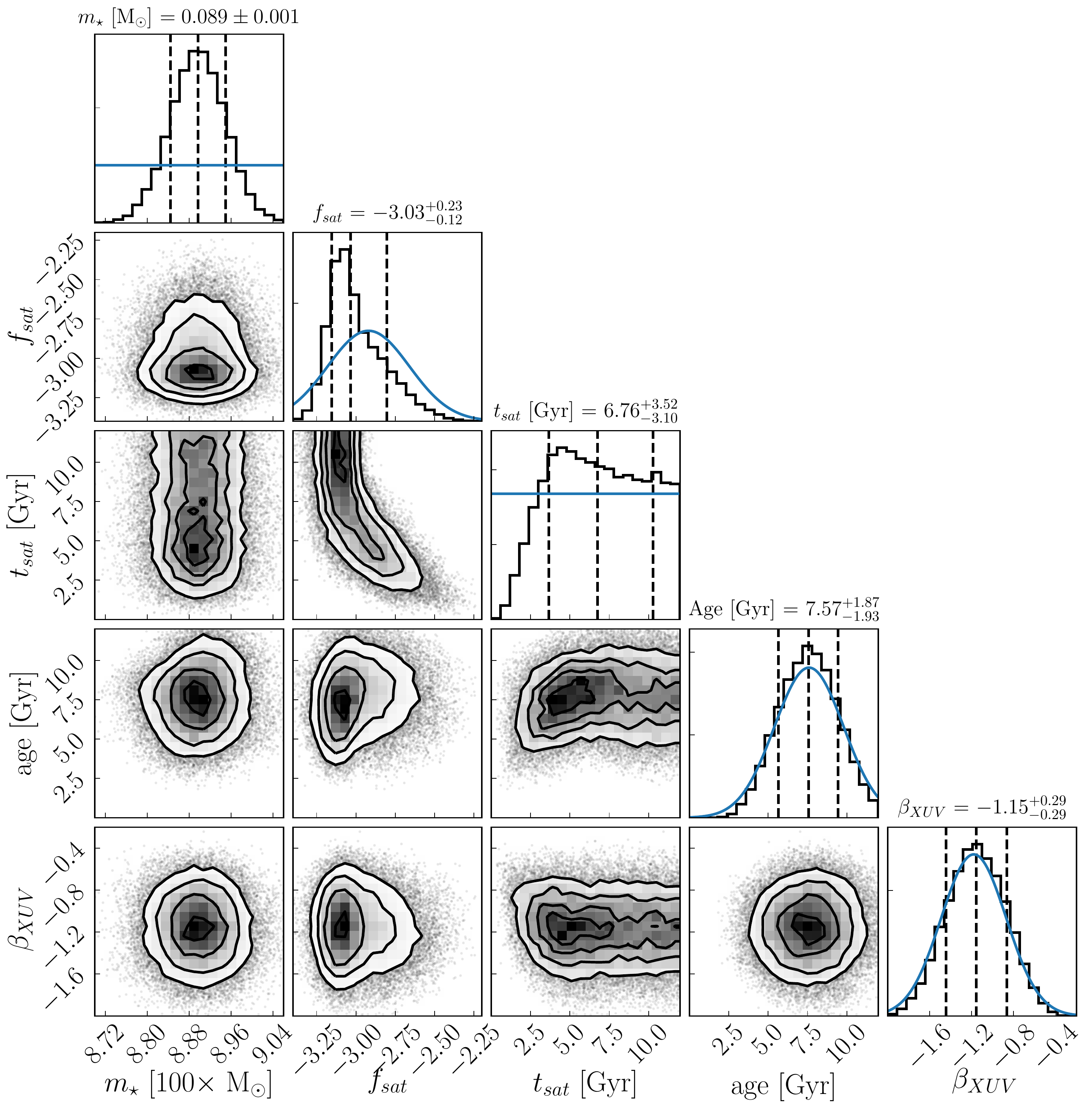}
   \caption{Same format as Fig.~\ref{fig:corner}, but derived by \approxposterior. \approxposterior recovered constraints and parameter correlations that are in good agreement with the \emcee MCMC, but requiring $980\times$ less computational resources.}%
    \label{fig:approx}%
\end{figure*}

We compare the approximate posterior distribution derived using \approxposterior with our previous results (referred to as the fiducial MCMC). We display the approximate joint and marginal posterior distributions in Fig.~\ref{fig:approx} and list the marginal constraints derived by both methods in Table~\ref{tab:constraints}.

\begin{figure*}
\centering
	\includegraphics[width=0.75\textwidth]{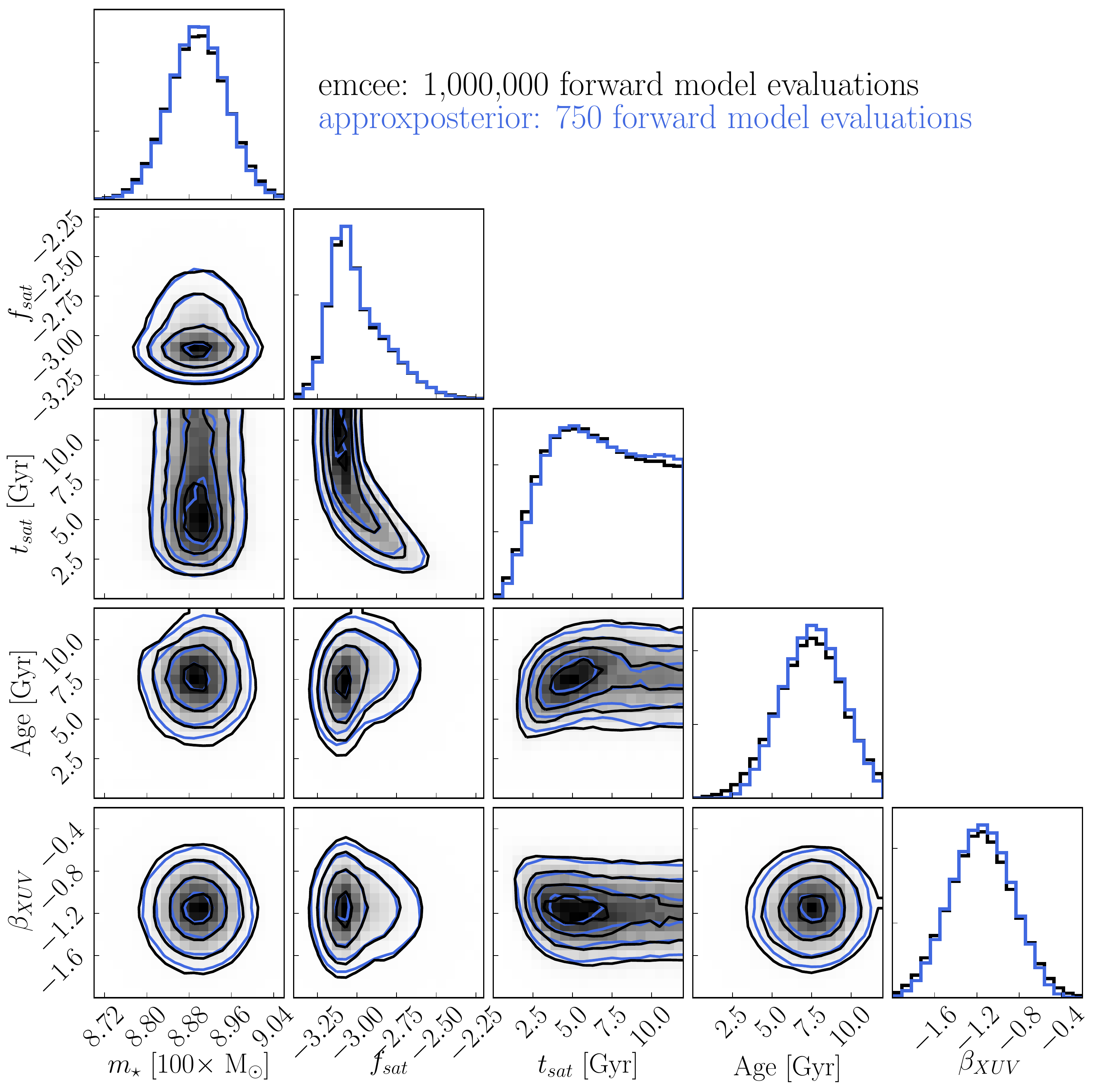}
   \caption{Same format as Fig.~\ref{fig:corner}, but with the fiducial posterior distribution in black and the \approxposterior-derived posterior distribution in blue. The joint and marginal posterior distributions estimated by \approxposterior are in excellent agreement with our fiducial \emcee-derived results.}%
    \label{fig:stacked}%
\end{figure*}

As seen in Fig.~\ref{fig:approx}, \approxposterior recovers the non-trivial correlations between model parameters seen in the fiducial MCMC posterior distribution. We emphasize this good agreement by overplotting the \approxposterior estimated posterior distribution (blue) on top of the fiducial MCMC results (black) in Fig.~\ref{fig:stacked}.

Our parameter constraints derived using \approxposterior are in good agreement with those inferred using \emcee. We find average errors in parameter medians and $1\sigma$ uncertainties of $0.61\%$ and $5.5\%$, respectively, relative to the constraints deriving using \emcee. These differences are larger than the MCSEs because the GP employed by \approxposterior is an accurate, yet imperfect, surrogate for the lnprobability calculation. \approxposterior tends to underestimate parameter uncertainties by a few percent because its algorithm preferentially selects high-likelihood points to expand its training set (see $\S$~\ref{app:augment}). This concentration of high-likelihood points slightly biases the inferred GP scale lengths, $l$, towards smaller values, effectively overfitting. The smaller values of $l$ shrink the estimated posterior distribution, producing the underestimated parameter uncertainties. We mitigate this effect by adding a small white noise term to the diagonal of the GP covariance matrix.

Not only can \approxposterior accurately recover Bayesian parameter constraints and correlations, it does so extremely quickly. \approxposterior requires only about 4 core hours to estimate the approximate posterior distribution, a factor of $980\times$ faster than our fiducial MCMC. Moreover, \approxposterior used $1330\times$ fewer \vplanet simulations to build its training set than the ${\sim}10^6$ simulations ran by the fiducial MCMC for likelihood evaluations. This reduction in computational expense arises from a combination of \approxposterior's GP-based lnprobability predictions only taking ${\sim}130\mu$s, compared to the much longer $10$s per \vplanet simulation, and its intelligent iterative training set augmentation algorithm. \approxposterior's efficient selection of the GP's training set focuses on high-likelihood regions to improve the GP's predictive ability in relevant regions of parameter space while minimizing the training set size.

Our findings demonstrate that \approxposterior can be used to estimate accurate approximations to the posterior probability distributions of the parameters that control stellar XUV evolution in late M dwarfs, but significantly faster than traditional MCMC methods. Note that \approxposterior is agnostic to the underlying forward model it learns on, enabling Bayesian parameter inference with other computationally-expensive forward models.


\begin{deluxetable*}{lcccc}
\caption{Parameter Constraints and Errors} \label{tab:constraints}
\tabletypesize{\small}
\tablehead{
\colhead{Parameter [units]} & \colhead{\vplanet-\emcee MCMC} & \colhead{\approxposterior MCMC} & \colhead{\approxposterior Relative Error} & \colhead{Monte Carlo Error}
}
\startdata
$m_\star$ [$M_{\odot}$] & $0.089^{+0.001}_{-0.001}$ &  $0.089^{+0.001}_{-0.001}$ & ${<}0.1\%$ & $3.41\times 10^{-6}$ \\  
$f_{sat}$ & $-3.03^{+0.23}_{-0.12}$ & $-3.03^{+0.23}_{-0.12}$ & ${<}0.1\%$ & $1.23\times 10^{-3}$  \\
$t_{sat}$ [Gyr] & $6.64^{+3.53}_{-3.13}$ & $6.76^{+3.52}_{-3.10}$ & $1.81\%$ & $2.0\times 10^{-2}$ \\
age [Gyr] & $7.46^{+2.01}_{-2.10}$ & $7.57^{+1.87}_{-1.93}$ & $1.47\%$ & $1.30 \times 10^{-3}$ \\
$\beta_{XUV}$ & $-1.16^{+0.31}_{-0.30}$ & $-1.15^{+0.29}_{-0.29}$ & $0.86\%$ & $2.12\times 10^{-3}$ \\
P$(\mathrm{saturated})$ & $0.40$ & $0.39$ & $2.5\%$ & $3.30 \times 10^{-3}$ \\
\enddata \vspace*{0.1in}
\tablecomments{Best fit values and uncertainties are derived using the medians, $16^{th}$, and $84^{th}$ percentiles from the marginal posterior distributions, respectively. P$(\mathrm{saturated})$ indicates the posterior probability that TRAPPIST-1 is still in the saturated regime today. The relative errors are computed as the absolute percent difference between the best fit values derived by \emcee and \approxposterior. The \approxposterior-derived results are in good agreement with the fiducial \emcee MCMC.}
\end{deluxetable*}

\subsection{TRAPPIST-1's Evolutionary History and Its Planets' XUV Environment} \label{sec:evol}

\begin{figure*}[t]
	\includegraphics[width=\textwidth]{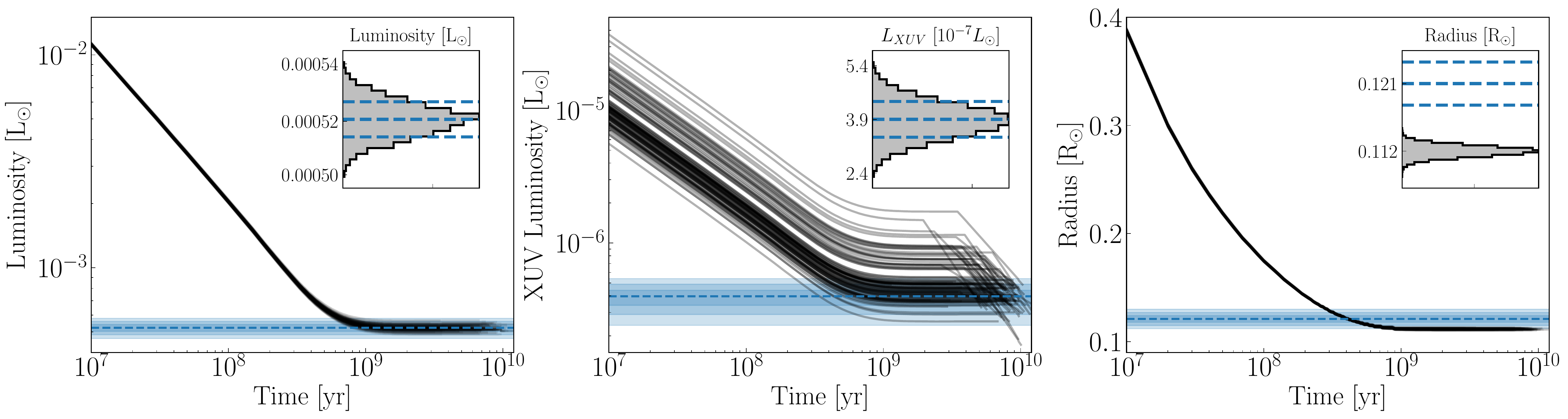}
   \caption{Plausible evolutionary histories of TRAPPIST-1's $L_{bol}$ (left), $L_{XUV}$ (center), and radius (right) using 100 samples drawn from the posterior distribution and simulated with \vplanet. In each panel, the blue shaded regions display the 1, 2, and 3 $\sigma$ uncertainties. The insets display the marginal distributions (black) evaluated at the age of the system, with the blue dashed lines indicating the observed value and +/- 1 $\sigma$ uncertainties. The radius, $L_{bol}$, and $L_{XUV}$ constraints are adopted from \citet{vanGrootel2018} and \citet{Wheatley2017}, respectively, by convolving the \citet{vanGrootel2018} $L_{bol}$ measurement with the $L_{XUV}/L_{bol}$ constraints from \citet{Wheatley2017}.}%
    \label{fig:evol}%
\end{figure*}

Here we consider plausible stellar evolutionary histories for TRAPPIST-1 by simulating 100 samples from the posterior distribution. We plot the evolution of TRAPPIST-1's $L_{bol}$, $L_{XUV}$, and radius in Fig.~\ref{fig:evol} and compare our models to the measured values. 

TRAPPIST-1 remains saturated throughout its $1$ Gyr-long pre-main sequence, with both $L_{XUV}$ and $L_{bol}$ decreasing by a factor of ${\sim}40$ before stabilizing on the main sequence. TRAPPIST-1's radius likely shrank by roughly a factor of 4 along the pre-main sequence. We derive a present-day radius $R_{\star} = 0.112 \pm{0.001} \ R_{\odot}$ from the posterior distribution, a value that is ${\sim} 7\%$ smaller than the \citet{vanGrootel2018} constraint, $R_{\star} = 0.121 \pm {0.003} \ R_{\odot}$, that was computed from their inferred mass and TRAPPIST-1's density \citep{Delrez2018}. This difference arises from the likely underprediction of TRAPPIST-1's radius by the \citet{Baraffe2015} models, consistent with stellar evolution models often underestimating the radii of late M dwarfs \citep{Reid2005,Spada2013}. 

An alternate explanation to account for its inflated radius is that TRAPPIST-1 has super-solar metallicity \citep{Burgasser2017,vanGrootel2018}, but \citet{vanGrootel2018} found in their modeling that TRAPPIST-1 required a metallicity of [Fe/H] = 0.4 to reproduce its density and radius. \citet{vanGrootel2018} show that as this result is $4.5\sigma$ off from the best fit value from \citet{Gillon2016}, who found [Fe/H] $= 0.04 \pm{0.08}$. The super-solar hypothesis is therefore strongly disfavored by the observational data. If we instead compute the radius from our marginal stellar mass posterior distribution and the observed density \citep{Delrez2018}, we obtain $R_{\star} = 0.120 \pm{0.002} \ R_{\odot}$, in agreement with \citet{vanGrootel2018} who used the same procedure.

\begin{figure}
	\includegraphics[width=0.98\columnwidth]{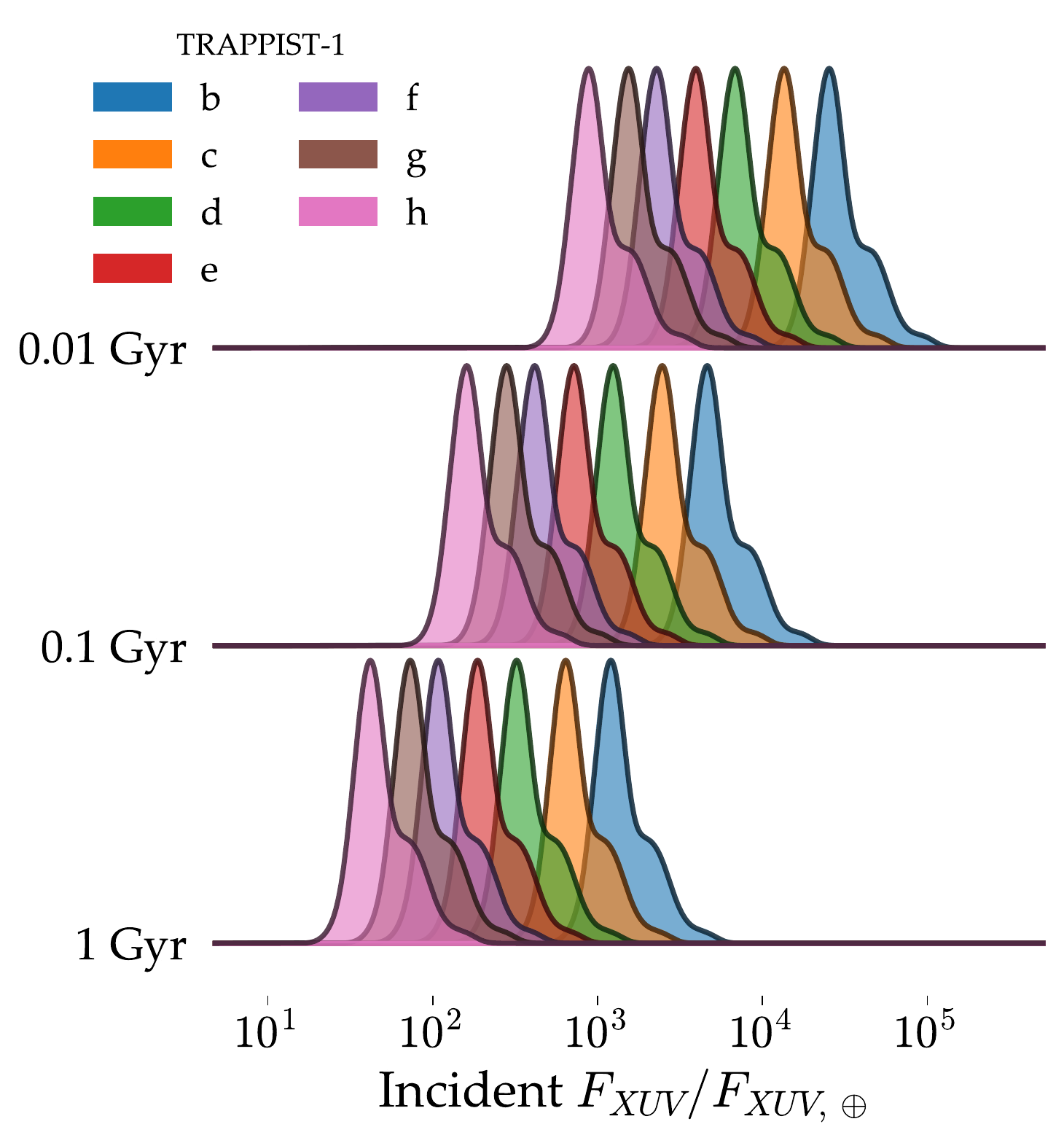}
   \caption{$F_{XUV}/F_{XUV,\oplus}$ for each TRAPPIST-1 planet derived from samples drawn from the posterior distribution and simulated using \vplanet when the system was 0.01, 0.1, and 1 Gyr old. The latter age corresponds to the approximate age at which TRAPPIST-1 entered the main sequence. The TRAPPIST-1 planetary system has likely endured a long-lasting extreme XUV environment.}%
    \label{fig:fluxes}%
\end{figure}

Since TRAPPIST-1 could still be saturated today, its planetary system has likely experienced a persistent extreme XUV environment. In Fig.~\ref{fig:fluxes}, we probe the distribution of XUV fluxes, $F_{XUV}$, derived from our posterior distributions for each TRAPPIST-1 planet when the system was 0.01, 0.1, and 1 Gyr old. We normalize these values by the $F_{XUV}$ received by Earth during the mean solar cycle \citep[$F_{XUV,\oplus} = 3.88$ erg s$^{-1}$cm$^{-2}$,][]{Ribas2005} and assume the planets remained near their current semi-major axes after migration in the natal protoplanetary disk halted \citep{Luger2017}. 

We infer that TRAPPIST-1b likely received extreme $F_{XUV}/F_{XUV, \oplus} \gsim 10^4$ during the early pre-main sequence before decaying to the present-day $F_{XUV}/F_{XUV, \oplus} \approx 10^3$, consistent with estimates from \citet{Wheatley2017}. The extended upper-tail of the $F_{XUV}$ distributions corresponds to the large $f_{sat}$ values permitted by the posterior distributions. The likely habitable zone planets, e, f, and g, similarly experienced severe XUV fluxes ranging from $F_{XUV}/F_{XUV, \oplus} \approx 10^2 - 10^{3.5}$ throughout the pre-main sequence. Even today, e, f, and g receive $F_{XUV}/F_{XUV, \oplus} \approx 10^2$, far in excess of the modern Earth, due to TRAPPIST-1's large present $L_{XUV}$, its extended saturated phase, and the close proximity of M dwarf HZ planets to their host star. These significant high energy fluxes likely drove an extended epoch of substantial atmospheric escape and water loss from the TRAPPIST-1 planets, potentially producing substantial abiotic O$_2$ atmospheres \citep{Luger2015,Bolmont2017,Bourrier2017a}.

\section{Discussion and Conclusions} \label{sec:discussion}

Here, we used MCMC to derive probabilistic constraints for TRAPPIST-1's stellar and $L_{XUV}$ evolution to characterize the evolving XUV environment of its planetary system. We inferred that TRAPPIST-1 likely maintained high $L_{XUV}/L_{bol} \approx 10^{-3}$ throughout its lifetime, with a $40\%$ chance that TRAPPIST-1 is still in the saturated regime today. Our results indicate that at least some ultracool dwarfs can sustain large $L_{XUV}$ in the saturated regime for Gyrs, consistent with activity lifetimes of late M dwarfs \citep{West2008}. We suggest that studies of volatile loss from planets orbiting ultracool dwarfs model the long-term $L_{XUV}$ evolution of the host star, or at least assume saturation timescales of $t_{sat}{\gsim}4$ Gyrs. Our choice of prior distributions strongly impacts our results as our inference hinges on only two measured properties of TRAPPIST-1, $L_{XUV}$ and $L_{bol}$. To mitigate this effect, we consulted previous studies and empirical observations of the activity evolution of late M dwarfs to construct realistic prior distributions.

The TRAPPIST-1 planets likely experienced significant XUV fluxes during the pre-main sequence, potentially driving extreme atmospheric erosion and water loss \citep{Bolmont2017,Bourrier2017a}. The high-energy fluxes incident on the inner-most planets throughout this phase were probably large enough for atmospheric mass loss to be recombination-limited ($F_{UV} \gsim 10^4$ g s$^{-1}$ cm$^{-2}$) and scale as $\dot{m} \sim F_{XUV}^{0.6}$ \citep{MurrayClay2009}, as opposed to the oft-assumed energy-limited escape \citep[$\dot{m} \sim F_{XUV}$,][]{Watson1981,Lammer2003}, potentially inhibiting volatile loss. If the TRAPPIST-1 planets did lose significant amounts of water as our estimates suggest, they must have formed with a large initial volatile inventory to account for their observed low densities \citep{Grimm2018}.

We demonstrated that the open source Python machine learning package, \approxposterior \citep{FlemingVanderPlas2018}, can efficiently compute an accurate approximation to the posterior distribution using an adaptive learning GP-based method, requiring $1330\times$ fewer \vplanet simulations and a factor of $980\times$ less core hours than traditional MCMC approaches. The posterior distributions derived by \approxposterior reproduced the non-trivial parameter correlations and best-fit values uncovered by our fiducial MCMC analysis. We find that \approxposterior recovers the best-fit values and $1\sigma$ uncertainties of our model parameters with an average error of $0.61\%$ and $5.5\%$, respectively, relative to our constraints derived using \emcee. If future observations of TRAPPIST-1 refine its fundamental parameters, and possibly $L_{XUV}/L_{bol}$, \approxposterior can be used to rapidly and accurately replicate our analysis to update our constraints.  

Finally, we note that our methodology constrains parameters that describe the long-term XUV evolution of TRAPPIST-1, conditioned on measurements. In principle, this approach can be extended to obtain evolutionary histories of planetary systems in general.  For example, in Figures~\ref{fig:evol} and \ref{fig:fluxes}, we examined the long-term evolution of TRAPPIST-1 and the evolving XUV fluxes received by its planetary system, respectively, with samples drawn from the posterior distribution. Future research can combine those results with additional physical effects, e.g. water loss or tidal dissipation, to build a probabilistic model for the long-term evolution of the planetary system, given our model for the underlying physics, to characterize its present state. In other words we could infer the evolutionary history of a planet or planetary system given suitable observational constraints. While simulating additional physical effects will inevitably increase the computational expense, we have demonstrated that \approxposterior can enable such efforts and provide insight into the histories of stars and their planets.


\acknowledgments
We thank the anonymous referee for their careful reading of our manuscript and insightful comments. This work was facilitated though the use of advanced computational, storage, and networking infrastructure provided by the Hyak supercomputer system and funded by the Student Technology Fund at the University of Washington. DPF was supported by NASA Headquarters under the NASA Earth and Space Science Fellowship Program - Grant 80NSSC17K0482.  This work was supported by the NASA Astrobiology Program Grant Number 80NSSC18K0829 and benefited from participation in the NASA Nexus for Exoplanet Systems Science research coordination network.

\software{\approxposterior: \citet{FlemingVanderPlas2018}, \texttt{corner}: \citet{ForemanMackey2016}, \texttt{emcee}: \citet{ForemanMackey2013}, \texttt{george}: \citet{george}, \vplanet: \citet{Barnes2019}} 

\appendix
\approxposterior is an implementation of the ``Bayesian Active Posterior Estimation" (BAPE) algorithm developed by \citet{Kandasamy2017}, but with several modifications to afford the user more control over the inference. Below, we qualitatively describe this algorithm, define parameters, and suggest typical values. We then discuss \approxposterior's convergence scheme.

\section{\approxposterior Algorithm and Convergence} \label{sec:app}

Qualitatively, the \approxposterior algorithm is as follows. First, assume a forward model with $d$ input parameters that is designed to reproduce some set of observations. In our case, $d$, the dimensionality of parameter space, is five. The model parameters have an input domain, $D$, that is defined by the user. The parameters are further described by a prior probability distribution based on the user's prior belief for how the model parameters are distributed.  Next, the user generates a training set, $T$, consisting of $m_0$ forward model simulations distributed across the parameter space. The user chooses how the $m_0$ samples are distributed throughout parameter space according to their preferred experimental design. \approxposterior then trains a GP on $T$ to construct a non-parametric model (sometimes called a ``surrogate model") that represents the outcomes of the forward model over the parameter space. Crucially, GPs also generate an uncertainty for the surrogate model at every point in parameter space.

\approxposterior then identifies $m$ more locations in parameter space to apply the forward model and add to $T$. The new locations are selected by determining the regions that the GP has identified as having both a high lnprobability, i.e. high posterior density, and a high predictive uncertainty. This selection is accomplished by maximizing a utility function ($u$, described below) that quantifies where the GP predicts high posterior density and high uncertainty in parameter space, focusing resources on parameter combinations that are likely to be consistent with the observations. \approxposterior re-trains the GP with the augmented $T$. The GP is then passed to an MCMC algorithm, e.g. \emcee, that samples the parameter space to obtain the approximate posterior distributions of the model parameters.

At the end of each iteration, \approxposterior checks if a convergence condition (described in $\S$~\ref{sec:app:convergence}) has been met. If the algorithm  has not yet converged, \approxposterior selects an additional $m$ new points to add to $T$, re-trains the GP, and again estimates the posterior distribution. This process repeats until convergence or until \approxposterior has run the maximum number of iterations, $n_{max}$, set by the user. In Algorithm~\ref{app:algo}, we list the aforementioned steps that comprise this algorithm.

\begin{algorithm} \label{app:algo}
\SetAlgoLined
 Assume an input domain $D$, GP prior on $f(\textbf{x})$ \\
 Generate a training set, $T$, consisting of $m_0$ pairs of $(\textbf{x}, f(\textbf{x}))$ \\
 \For{$t=0, 1, ..., n_{\mathrm{max}}$}{
    \For{$i=0, 1, ..., m$}{
      Find \textbf{x}$^+$ = argmax$_{\textbf{x} \in D}$ $u(\textbf{x})$ \\
       Compute $f(\textbf{x$^+$})$ \\
       Append $(\textbf{x$^+$}, f(\textbf{x$^+$}))$ to $T$ \\
       Re-train GP, optimize GP hyperparameters given augmented $T$ \\
   }
   Use MCMC to obtain approximate posterior distribution with GP surrogate for $f(\textbf{x})$ \\
   \If{$\mathrm{converged}$}{
        \textbf{break} \\
    }
 }
\caption{\approxposterior Approximate Inference Pseudo Code}
\end{algorithm}

In Algorithm~\ref{app:algo}, we define $f(\textbf{x}) = \mathcal{\ln L}(\textbf{x})$ + $\ln \mathrm{Prior}(\textbf{x})$ as the lnprobability function used for MCMC sampling with \emcee and \textbf{x}$^+$ as the point in parameter space selected by maximizing $u$. For our application, evaluating $f(\textbf{x})$ requires running a \vplanet simulation to compute $\mathcal{\ln L}(\textbf{x})$ (see $\S$~\ref{sec:mcmc:like}). By placing a GP prior with a squared exponential kernel on $f(\textbf{x})$, we assume that the function is smooth and continuous, both reasonable assumptions for modeling the posterior density. For inference problems that are liable to violate these assumptions, other kernels, e.g. the Ornstein-Uhlenbeck kernel, may be more appropriate (we refer the reader to \citet{Rasmussen2006} for detailed descriptions of common GP kernels and their mathematical properties). \approxposterior uses \texttt{george} \citep{george} for all GP calculations and hence users can apply any kernels implemented in that software package. 

\approxposterior has several free parameters that can be set by the user: $m_0$, the size of the initial training set (50 in our case), $n_{\mathrm{max}}$, the maximum number of iterations (15), $m$, the number of new points to select each iteration where the forward model will be evaluated (100 per iteration), and $\epsilon$, the convergence threshold (0.1). Typically, we find that $n_{\mathrm{max}}=2-3 \times d$, $m, m_0 = 10-20 \times d$, and $\epsilon = 0.1$ work well in practice, although performance may vary depending on the use case. For a complete list of \approxposterior parameters, we refer the reader to the online documentation.\footnote{ \href{https://dflemin3.github.io/approxposterior}{https://dflemin3.github.io/approxposterior/}}

Note that \approxposterior does not linearly transform the parameter space to the unit hypercube as did \citet{Kandasamy2017}. Moreover, \approxposterior does not fix the covariance scale lengths, instead opting to estimate all GP kernel hyperparameters by maximizing the marginal likelihood of the GP, given its training set, at a user-specified cadence. In Algorithm~\ref{app:algo}, we optimize the GP hyperparameters each time a new point is added to the training set, but in practice we found this is unnecessary, especially at later iterations when the GP has developed a reasonable approximation of the posterior. The authors prefer to optimize the GP hyperparameters twice per iteration, once after half of the $m$ new points have been selected, and again after all $m$ points have been selected.

\subsection{Augmenting the Training Set} \label{app:augment}

Each iteration, \approxposterior selects $m$ new points to add to the GP's training set by maximizing the utility function, $u$. To motivate the choice of $u$, consider the following argument based on \citet{Kandasamy2017}: \approxposterior assumes that the forward model the GP learns on, here \vplanet via $\ln \mathcal{L}$, is computationally-expensive to run, and hence \approxposterior seeks to minimize the number of forward model evaluations required to build its training set. For inference problems, it is natural to select high-lnprobability regions in parameter space to augment the GP training set as this is where the posterior density is large. Furthermore, selecting regions in parameter space where the GP's predictive uncertainty is already small offers little value, compared to regions where its predictions are more uncertain, as additional points in low-uncertainty regions are unlikely to alter the GP's predictions.

With these considerations in mind, \citet{Kandasamy2017} leverage the analytic properties of GPs to derive the ``exponentiated variance" utility function, given by their Eq.~(5)
\begin{equation} \label{app:eq:bape}
    u_{\textrm{EV}}(\textbf{x}) = \exp(2 \mu_t(\textbf{x}) + \sigma_t^2(\textbf{x}))(\exp(\sigma_t^2(\textbf{x})) - 1),
\end{equation}
where $\mu_t(\textbf{x})$ and $\sigma_t^2(\textbf{x})$ are the mean and variance of the GP's predictive conditional distribution evaluated at \textbf{x}, respectively, for the $t^{th}$ \approxposterior iteration. To select each point, we maximize Eqn.~(\ref{app:eq:bape}) using the Nelder-Mead method \citep{Nelder1965}. Note that this optimization is rather cheap since it only requires evaluating the GP's predictive conditional distribution, so this task is not a significant computational bottleneck. We restart this optimization 5 times to reduce the influence of local extrema. Note that in practice, we optimize the natural logarithm of the utility function to ensure numerical stability.

\begin{figure*}[h]
\centering
	\includegraphics[width=0.75\textwidth]{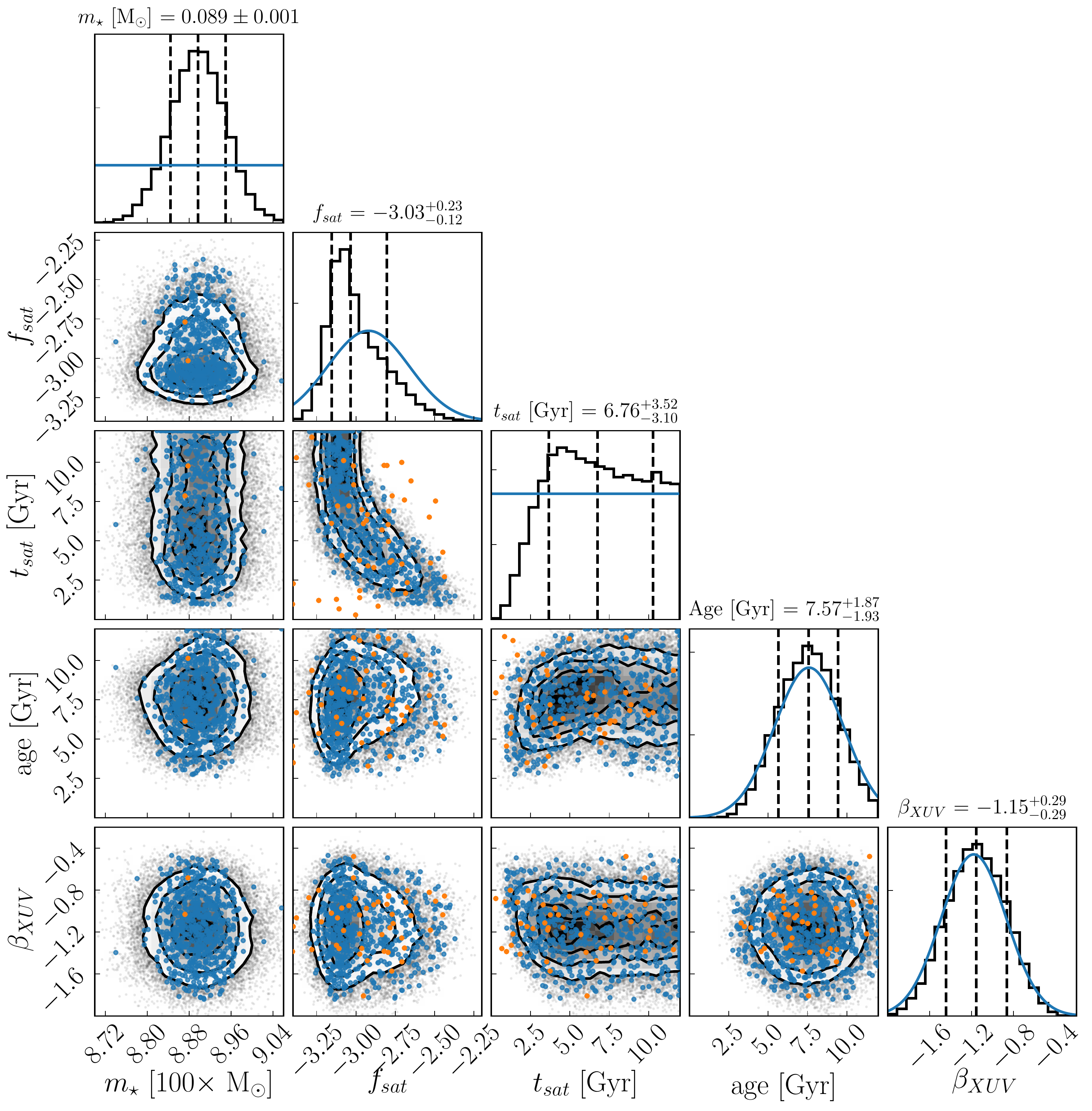}
   \caption{Same as Fig.~\ref{fig:approx}, but overplotted with the training set for \approxposterior's GP. The orange points display the initial training points whereas the blue points display the points iteratively selected by maximizing the \citet{Kandasamy2017} utility function, Eqn.~(\ref{app:eq:bape}). By design, \approxposterior selected points to expand its training set in regions of high posterior density, improving its GP's predictive accuracy in the most relevant regions of parameter space while seldom wasting computational resources in the low likelihood regions.}%
    \label{fig:points}%
\end{figure*}

As demonstrated in \citet{Kandasamy2017}, Eqn.~(\ref{app:eq:bape}) identifies high-likelihood points where the GP's predictions are uncertain, significantly reducing the cost of training an accurate GP surrogate model. We highlight this behavior for our own application in Fig.~\ref{fig:points} by displaying the approximate posterior distribution derived by \approxposterior from Fig.~\ref{fig:approx} overplotted with the initial training set in orange and the points selected by sequentially maximizing Eqn.~(\ref{app:eq:bape}) in blue. Given the small initial training set, \approxposterior successfully selects high-posterior density points in parameter space to augment the GP's training set. Some points are selected in low-likelihood regions early on, typically near the edges of parameter space where the GP's uncertainty was initially large.


\subsection{Convergence} \label{sec:app:convergence}

We assess the convergence of the \approxposterior algorithm by comparing the means of the approximate marginal posterior distributions over successive iterations. We consider an \approxposterior run ``converged" if the differences between the marginal posterior means, relative to the widths of the marginal posteriors, are less than a tolerance parameter, $\epsilon$, for $k_{max}$ consecutive iterations. Effectively, this criterion checks if the expected value of each model parameter over the posterior distribution varies by ${\leq}{\epsilon}$ standard deviations from the previous iteration's expected values. That is, we require the \approxposterior convergence diagnostic $z_{t,j}{\leq}{\epsilon}$ for all $j$, where
\begin{equation}
    z_{t,j} = |\mu_{t,j} - \mu_{t-1,j}| / \sigma_{t-1,j},
\end{equation}
 and $\mu_{t,j}$ and $\sigma_{t,j}$ are the mean and standard deviation of the approximate marginal posterior distribution for the $t^{th}$ iteration and the $j^{th}$ parameter. This quantity is analogous to the ``z-score" commonly used in many statistical tests. Following \citet{Wang2018}, we require this condition to be satisfied for $k_{max}$ consecutive iterations to ensure \approxposterior is producing a consistent result. With this scheme, \approxposterior tolerates deviations from the previous estimate that are less than, or at least consistent with, the previous values, given the inherent uncertainty implied by the width of the posterior distribution. For our application, we adopted conservative choices of $\epsilon = 0.1$ and $k_{max} = 5$. Each \approxposterior iteration, we also visually inspected the estimated posterior distribution to ensure convergence. 

\begin{figure*}[h]
\centering
	\includegraphics[width=0.75\textwidth]{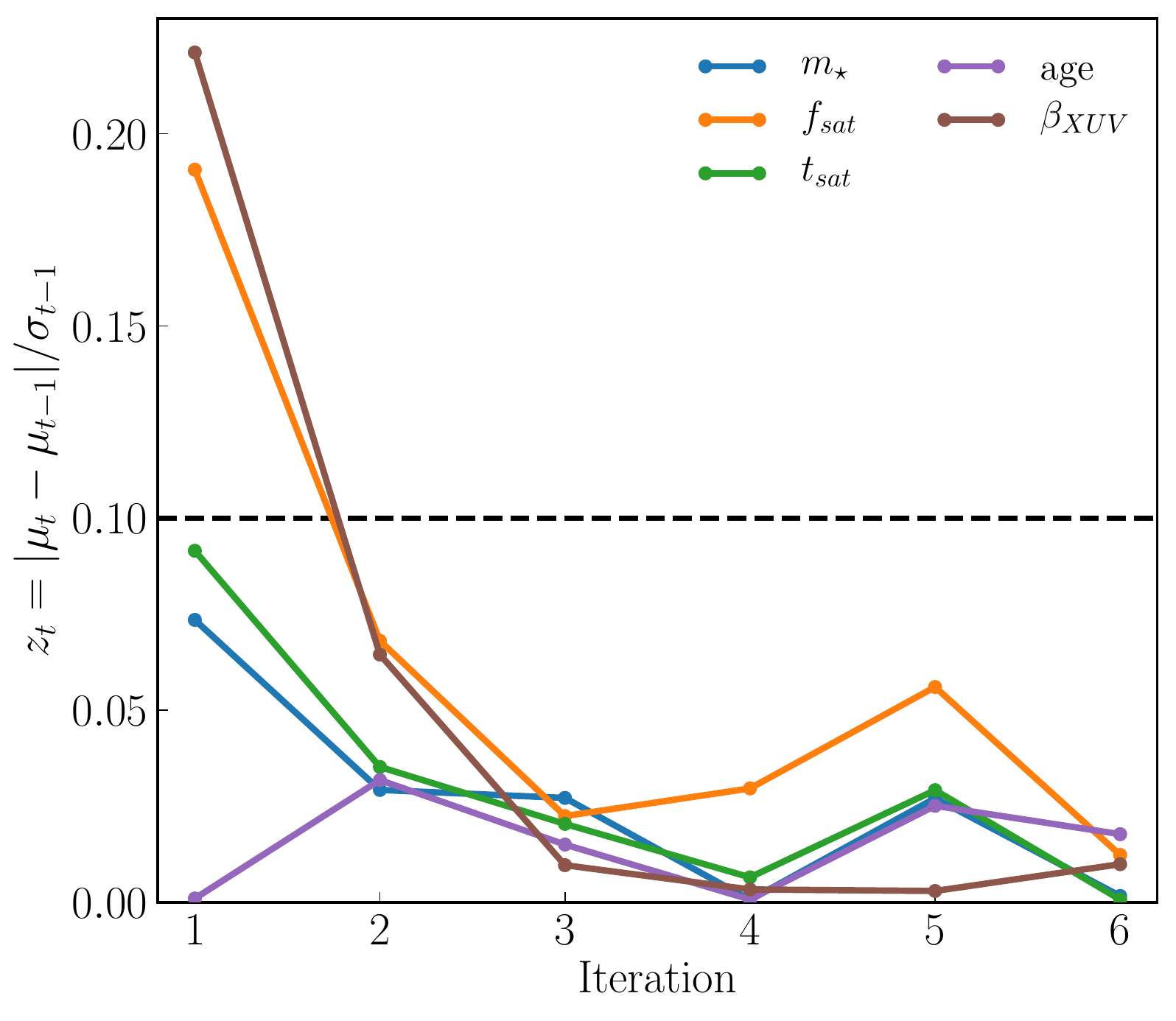}
   \caption{The \approxposterior convergence diagnostic, $z_t$, as a function of iteration for the run presented in the main text. Note that in \approxposterior, the initial iteration is iteration 0. The black dashed line indicates our adopted convergence threshold of $\epsilon = 0.1$. \approxposterior quickly converges to a consistent and accurate result.}%
    \label{fig:convergence}%
\end{figure*}

In Fig.~\ref{fig:convergence}, we display the convergence diagnostic quantity, $z_t$, as a function of iteration for each model parameter for the \approxposterior run presented in the main text. \approxposterior quickly finds a consistent result as $z_t$ decreases below our convergence threshold within the first few iterations. For each parameter, $z_t$ continues to decrease until iteration 3 before stabilizing. The evolution of $z_t$ is not monotonic, however, owing to the stochastic nature of GPs, our hyperparameter optimization scheme, and MCMC sampling that can cause these values to occasionally be slightly worse than previous iterations. Requiring convergence over $k_{max}$ consecutive iterations mitigates the impact of this stochasticity.

\bibliography{trappist}

\begin{thebibliography}{}
\expandafter\ifx\csname natexlab\endcsname\relax\def\natexlab#1{#1}\fi
\providecommand{\url}[1]{\href{#1}{#1}}

\bibitem[{{Ambikasaran} {et~al.}(2014){Ambikasaran}, {Foreman-Mackey},
  {Greengard}, {Hogg}, \& {O'Neil}}]{george}
{Ambikasaran}, S., {Foreman-Mackey}, D., {Greengard}, L., {Hogg}, D.~W., \&
  {O'Neil}, M. 2014

\bibitem[{{Baraffe} {et~al.}(2015){Baraffe}, {Homeier}, {Allard}, \&
  {Chabrier}}]{Baraffe2015}
{Baraffe}, I., {Homeier}, D., {Allard}, F., \& {Chabrier}, G. 2015, \aap, 577,
  A42

\bibitem[{{Barclay} {et~al.}(2018){Barclay}, {Pepper}, \&
  {Quintana}}]{Barclay2018}
{Barclay}, T., {Pepper}, J., \& {Quintana}, E.~V. 2018, \apjs, 239, 2

\bibitem[{{Barnes} {et~al.}(2014){Barnes}, {Jenkins}, {Jones}, {Jeffers},
  {Rojo}, {Arriagada}, {Jord{\'a}n}, {Minniti}, {Tuomi}, {Pinfield}, \&
  {Anglada-Escud{\'e}}}]{Barnes2014}
{Barnes}, J.~R., {Jenkins}, J.~S., {Jones}, H.~R.~A., {et~al.} 2014, \mnras,
  439, 3094

\bibitem[{{Barnes} {et~al.}(2019){Barnes}, {Luger}, {Deitrick}, {Driscoll},
  {Quinn}, {Fleming}, {Smotherman}, {McDonald}, {Wilhelm}, {Garcia}, {Barth},
  {Guyer}, {Meadows}, {Bitz}, {Gupta}, {Domagal-Goldman}, \&
  {Armstrong}}]{Barnes2019}
{Barnes}, R., {Luger}, R., {Deitrick}, R., {et~al.} 2019, arXiv e-prints,
  arXiv:1905.06367

\bibitem[{{Bird} {et~al.}(2019){Bird}, {Rogers}, {Peiris}, {Verde},
  {Font-Ribera}, \& {Pontzen}}]{Bird2019}
{Bird}, S., {Rogers}, K.~K., {Peiris}, H.~V., {et~al.} 2019, \jcap, 2, 050

\bibitem[{{Bolmont} {et~al.}(2017){Bolmont}, {Gallet}, {Mathis}, {Charbonnel},
  {Amard}, \& {Alibert}}]{Bolmont2017}
{Bolmont}, E., {Gallet}, F., {Mathis}, S., {et~al.} 2017, \aap, 604, A113

\bibitem[{{Bourrier} {et~al.}(2017{\natexlab{a}}){Bourrier}, {Ehrenreich},
  {Wheatley}, {Bolmont}, {Gillon}, {de Wit}, {Burgasser}, {Jehin}, {Queloz}, \&
  {Triaud}}]{Bourrier2017b}
{Bourrier}, V., {Ehrenreich}, D., {Wheatley}, P.~J., {et~al.}
  2017{\natexlab{a}}, \aap, 599, L3

\bibitem[{{Bourrier} {et~al.}(2017{\natexlab{b}}){Bourrier}, {de Wit},
  {Bolmont}, {Stamenkovi{\'c}}, {Wheatley}, {Burgasser}, {Delrez}, {Demory},
  {Ehrenreich}, {Gillon}, {Jehin}, {Leconte}, {Lederer}, {Lewis}, {Triaud}, \&
  {Van Grootel}}]{Bourrier2017a}
{Bourrier}, V., {de Wit}, J., {Bolmont}, E., {et~al.} 2017{\natexlab{b}}, \aj,
  154, 121

\bibitem[{{Burgasser} \& {Mamajek}(2017)}]{Burgasser2017}
{Burgasser}, A.~J., \& {Mamajek}, E.~E. 2017, \apj, 845, 110

\bibitem[{{Chadney} {et~al.}(2015){Chadney}, {Galand}, {Unruh}, {Koskinen}, \&
  {Sanz-Forcada}}]{Chadney2015}
{Chadney}, J.~M., {Galand}, M., {Unruh}, Y.~C., {Koskinen}, T.~T., \&
  {Sanz-Forcada}, J. 2015, \icarus, 250, 357

\bibitem[{Delfosse {et~al.}(1998)Delfosse, Forveille, Perrier, \&
  Mayor}]{Delfosse1998}
Delfosse, X., Forveille, T., Perrier, C., \& Mayor, M. 1998, Astronomy and
  Astrophysics, 331, 581

\bibitem[{{Delrez} {et~al.}(2018){Delrez}, {Gillon}, {Triaud}, {Demory}, {de
  Wit}, {Ingalls}, {Agol}, {Bolmont}, {Burdanov}, {Burgasser}, {Carey},
  {Jehin}, {Leconte}, {Lederer}, {Queloz}, {Selsis}, \& {Van
  Grootel}}]{Delrez2018}
{Delrez}, L., {Gillon}, M., {Triaud}, A.~H.~M.~J., {et~al.} 2018, \mnras, 475,
  3577

\bibitem[{{Dressing} \& {Charbonneau}(2015)}]{Dressing2015}
{Dressing}, C.~D., \& {Charbonneau}, D. 2015, \apj, 807, 45

\bibitem[{Flegal {et~al.}(2008)Flegal, Haran, \& Jones}]{Flegal2008}
Flegal, J.~M., Haran, M., \& Jones, G.~L. 2008, Statist. Sci., 23, 250.
\newblock \url{https://doi.org/10.1214/08-STS257}

\bibitem[{Flegal \& Jones(2010)}]{Flegal2010}
Flegal, J.~M., \& Jones, G.~L. 2010, Ann. Statist., 38, 1034.
\newblock \url{https://doi.org/10.1214/09-AOS735}

\bibitem[{{Fleming} \& {VanderPlas}(2018)}]{FlemingVanderPlas2018}
{Fleming}, D.~P., \& {VanderPlas}, J. 2018, The Journal of Open Source
  Software, 3, 781

\bibitem[{{Foreman-Mackey}(2016)}]{ForemanMackey2016}
{Foreman-Mackey}, D. 2016, The Journal of Open Source Software, 1,
  doi:10.21105/joss.00024

\bibitem[{{Foreman-Mackey} {et~al.}(2013){Foreman-Mackey}, {Hogg}, {Lang}, \&
  {Goodman}}]{ForemanMackey2013}
{Foreman-Mackey}, D., {Hogg}, D.~W., {Lang}, D., \& {Goodman}, J. 2013, \pasp,
  125, 306

\bibitem[{{Gillon} {et~al.}(2016){Gillon}, {Jehin}, {Lederer}, {Delrez}, {de
  Wit}, {Burdanov}, {Van Grootel}, {Burgasser}, {Triaud}, {Opitom}, {Demory},
  {Sahu}, {Bardalez Gagliuffi}, {Magain}, \& {Queloz}}]{Gillon2016}
{Gillon}, M., {Jehin}, E., {Lederer}, S.~M., {et~al.} 2016, \nat, 533, 221

\bibitem[{{Gillon} {et~al.}(2017){Gillon}, {Triaud}, {Demory}, {Jehin}, {Agol},
  {Deck}, {Lederer}, {de Wit}, {Burdanov}, {Ingalls}, {Bolmont}, {Leconte},
  {Raymond}, {Selsis}, {Turbet}, {Barkaoui}, {Burgasser}, {Burleigh}, {Carey},
  {Chaushev}, {Copperwheat}, {Delrez}, {Fernandes}, {Holdsworth}, {Kotze}, {Van
  Grootel}, {Almleaky}, {Benkhaldoun}, {Magain}, \& {Queloz}}]{Gillon2017}
{Gillon}, M., {Triaud}, A.~H.~M.~J., {Demory}, B.-O., {et~al.} 2017, \nat, 542,
  456

\bibitem[{{Gonzales} {et~al.}(2019){Gonzales}, {Faherty}, {Gagn{\'e}}, {Teske},
  {McWilliam}, \& {Cruz}}]{Gonzales2019}
{Gonzales}, E.~C., {Faherty}, J.~K., {Gagn{\'e}}, J., {et~al.} 2019, \apj, 886,
  131

\bibitem[{{Grimm} {et~al.}(2018){Grimm}, {Demory}, {Gillon}, {Dorn}, {Agol},
  {Burdanov}, {Delrez}, {Sestovic}, {Triaud}, {Turbet}, {Bolmont}, {Caldas},
  {de Wit}, {Jehin}, {Leconte}, {Raymond}, {Van Grootel}, {Burgasser}, {Carey},
  {Fabrycky}, {Heng}, {Hernandez}, {Ingalls}, {Lederer}, {Selsis}, \&
  {Queloz}}]{Grimm2018}
{Grimm}, S.~L., {Demory}, B.-O., {Gillon}, M., {et~al.} 2018, \aap, 613, A68

\bibitem[{{Jackson} {et~al.}(2012){Jackson}, {Davis}, \&
  {Wheatley}}]{Jackson2012}
{Jackson}, A.~P., {Davis}, T.~A., \& {Wheatley}, P.~J. 2012, \mnras, 422, 2024

\bibitem[{Kandasamy {et~al.}(2017)Kandasamy, Schneider, \&
  Póczos}]{Kandasamy2017}
Kandasamy, K., Schneider, J., \& Póczos, B. 2017, Artificial Intelligence,
  243, 45 .
\newblock
  \url{http://www.sciencedirect.com/science/article/pii/S0004370216301394}

\bibitem[{{Lammer} {et~al.}(2003){Lammer}, {Selsis}, {Ribas}, {Guinan},
  {Bauer}, \& {Weiss}}]{Lammer2003}
{Lammer}, H., {Selsis}, F., {Ribas}, I., {et~al.} 2003, \apjl, 598, L121

\bibitem[{{Lincowski} {et~al.}(2018){Lincowski}, {Meadows}, {Crisp},
  {Robinson}, {Luger}, {Lustig-Yaeger}, \& {Arney}}]{Lincowski2018}
{Lincowski}, A.~P., {Meadows}, V.~S., {Crisp}, D., {et~al.} 2018, \apj, 867, 76

\bibitem[{{Luger} \& {Barnes}(2015)}]{Luger2015}
{Luger}, R., \& {Barnes}, R. 2015, Astrobiology, 15, 119

\bibitem[{{Luger} {et~al.}(2017){Luger}, {Sestovic}, {Kruse}, {Grimm},
  {Demory}, {Agol}, {Bolmont}, {Fabrycky}, {Fernandes}, {Van Grootel},
  {Burgasser}, {Gillon}, {Ingalls}, {Jehin}, {Raymond}, {Selsis}, {Triaud},
  {Barclay}, {Barentsen}, {Howell}, {Delrez}, {de Wit}, {Foreman-Mackey},
  {Holdsworth}, {Leconte}, {Lederer}, {Turbet}, {Almleaky}, {Benkhaldoun},
  {Magain}, {Morris}, {Heng}, \& {Queloz}}]{Luger2017}
{Luger}, R., {Sestovic}, M., {Kruse}, E., {et~al.} 2017, Nature Astronomy, 1,
  0129

\bibitem[{{Lustig-Yaeger} {et~al.}(2019){Lustig-Yaeger}, {Meadows}, \&
  {Lincowski}}]{Lustig2019}
{Lustig-Yaeger}, J., {Meadows}, V.~S., \& {Lincowski}, A.~P. 2019, arXiv
  e-prints, arXiv:1905.07070

\bibitem[{{McClintock} \& {Rozo}(2019)}]{McClintock2019}
{McClintock}, T., \& {Rozo}, E. 2019, arXiv e-prints, arXiv:1905.09299

\bibitem[{{Morley} {et~al.}(2017){Morley}, {Kreidberg}, {Rustamkulov},
  {Robinson}, \& {Fortney}}]{Morley2017}
{Morley}, C.~V., {Kreidberg}, L., {Rustamkulov}, Z., {Robinson}, T., \&
  {Fortney}, J.~J. 2017, \apj, 850, 121

\bibitem[{{Morris} {et~al.}(2018){Morris}, {Agol}, {Davenport}, \&
  {Hawley}}]{Morris2018}
{Morris}, B.~M., {Agol}, E., {Davenport}, J.~R.~A., \& {Hawley}, S.~L. 2018,
  \apj, 857, 39

\bibitem[{{Murray-Clay} {et~al.}(2009){Murray-Clay}, {Chiang}, \&
  {Murray}}]{MurrayClay2009}
{Murray-Clay}, R.~A., {Chiang}, E.~I., \& {Murray}, N. 2009, \apj, 693, 23

\bibitem[{Nelder \& Mead(1965)}]{Nelder1965}
Nelder, J.~A., \& Mead, R. 1965, The Computer Journal, 7, 308.
\newblock \url{https://doi.org/10.1093/comjnl/7.4.308}

\bibitem[{{Parker}(1955)}]{Parker1955}
{Parker}, E.~N. 1955, \apj, 122, 293

\bibitem[{{Peacock} {et~al.}(2019){Peacock}, {Barman}, {Shkolnik},
  {Hauschildt}, \& {Baron}}]{Peacock2019}
{Peacock}, S., {Barman}, T., {Shkolnik}, E.~L., {Hauschildt}, P.~H., \&
  {Baron}, E. 2019, \apj, 871, 235

\bibitem[{{Pizzolato} {et~al.}(2003){Pizzolato}, {Maggio}, {Micela},
  {Sciortino}, \& {Ventura}}]{Pizzolato2003}
{Pizzolato}, N., {Maggio}, A., {Micela}, G., {Sciortino}, S., \& {Ventura}, P.
  2003, \aap, 397, 147

\bibitem[{Powell(1964)}]{Powell1964}
Powell, M. J.~D. 1964, The Computer Journal, 7, 155.
\newblock \url{https://doi.org/10.1093/comjnl/7.2.155}

\bibitem[{{Rasmussen} \& {Williams}(2006)}]{Rasmussen2006}
{Rasmussen}, C.~E., \& {Williams}, C.~K.~I. 2006, {Gaussian Processes for
  Machine Learning}

\bibitem[{{Reid} \& {Hawley}(2005)}]{Reid2005}
{Reid}, I.~N., \& {Hawley}, S.~L. 2005, {New light on dark stars : red dwarfs,
  low-mass stars, brown dwarfs}, doi:10.1007/3-540-27610-6

\bibitem[{{Ribas} {et~al.}(2005){Ribas}, {Guinan}, {G{\"u}del}, \&
  {Audard}}]{Ribas2005}
{Ribas}, I., {Guinan}, E.~F., {G{\"u}del}, M., \& {Audard}, M. 2005, \apj, 622,
  680

\bibitem[{{Roettenbacher} \& {Kane}(2017)}]{Roettenbacher2017}
{Roettenbacher}, R.~M., \& {Kane}, S.~R. 2017, \apj, 851, 77

\bibitem[{{Schneider} \& {Shkolnik}(2018)}]{Schneider2018}
{Schneider}, A.~C., \& {Shkolnik}, E.~L. 2018, \aj, 155, 122

\bibitem[{{Segura} {et~al.}(2005){Segura}, {Kasting}, {Meadows}, {Cohen},
  {Scalo}, {Crisp}, {Butler}, \& {Tinetti}}]{Segura2005}
{Segura}, A., {Kasting}, J.~F., {Meadows}, V., {et~al.} 2005, Astrobiology, 5,
  706

\bibitem[{{Shkolnik} \& {Barman}(2014)}]{Shkolnik2014}
{Shkolnik}, E.~L., \& {Barman}, T.~S. 2014, \aj, 148, 64

\bibitem[{{Skumanich}(1972)}]{Skumanich1972}
{Skumanich}, A. 1972, \apj, 171, 565

\bibitem[{{Spada} {et~al.}(2013){Spada}, {Demarque}, {Kim}, \&
  {Sills}}]{Spada2013}
{Spada}, F., {Demarque}, P., {Kim}, Y.-C., \& {Sills}, A. 2013, \apj, 776, 87

\bibitem[{{Vaiana} {et~al.}(1981){Vaiana}, {Cassinelli}, {Fabbiano},
  {Giacconi}, {Golub}, {Gorenstein}, {Haisch}, {Harnden}, {Johnson}, {Linsky},
  {Maxson}, {Mewe}, {Rosner}, {Seward}, {Topka}, \& {Zwaan}}]{Vaiana1981}
{Vaiana}, G.~S., {Cassinelli}, J.~P., {Fabbiano}, G., {et~al.} 1981, \apj, 245,
  163

\bibitem[{{Van Grootel} {et~al.}(2018){Van Grootel}, {Fernandes}, {Gillon},
  {Jehin}, {Manfroid}, {Scuflaire}, {Burgasser}, {Barkaoui}, {Benkhaldoun},
  {Burdanov}, {Delrez}, {Demory}, {de Wit}, {Queloz}, \&
  {Triaud}}]{vanGrootel2018}
{Van Grootel}, V., {Fernandes}, C.~S., {Gillon}, M., {et~al.} 2018, \apj, 853,
  30

\bibitem[{Wang \& Li(2018)}]{Wang2018}
Wang, H., \& Li, J. 2018, Neural Computation, 30, 3072.
\newblock \url{https://doi.org/10.1162/neco_a_01127}

\bibitem[{{Watson} {et~al.}(1981){Watson}, {Donahue}, \& {Walker}}]{Watson1981}
{Watson}, A.~J., {Donahue}, T.~M., \& {Walker}, J.~C.~G. 1981, \icarus, 48, 150

\bibitem[{{West} {et~al.}(2008){West}, {Hawley}, {Bochanski}, {Covey}, {Reid},
  {Dhital}, {Hilton}, \& {Masuda}}]{West2008}
{West}, A.~A., {Hawley}, S.~L., {Bochanski}, J.~J., {et~al.} 2008, \aj, 135,
  785

\bibitem[{{West} {et~al.}(2015){West}, {Weisenburger}, {Irwin},
  {Berta-Thompson}, {Charbonneau}, {Dittmann}, \& {Pineda}}]{West2015}
{West}, A.~A., {Weisenburger}, K.~L., {Irwin}, J., {et~al.} 2015, \apj, 812, 3

\bibitem[{{Wheatley} {et~al.}(2017){Wheatley}, {Louden}, {Bourrier},
  {Ehrenreich}, \& {Gillon}}]{Wheatley2017}
{Wheatley}, P.~J., {Louden}, T., {Bourrier}, V., {Ehrenreich}, D., \& {Gillon},
  M. 2017, \mnras, 465, L74

\bibitem[{{Wright} \& {Drake}(2016)}]{Wright2016}
{Wright}, N.~J., \& {Drake}, J.~J. 2016, \nat, 535, 526

\bibitem[{{Wright} {et~al.}(2011){Wright}, {Drake}, {Mamajek}, \&
  {Henry}}]{Wright2011}
{Wright}, N.~J., {Drake}, J.~J., {Mamajek}, E.~E., \& {Henry}, G.~W. 2011,
  \apj, 743, 48

\bibitem[{{Wright} {et~al.}(2018){Wright}, {Newton}, {Williams}, {Drake}, \&
  {Yadav}}]{Wright2018}
{Wright}, N.~J., {Newton}, E.~R., {Williams}, P.~K.~G., {Drake}, J.~J., \&
  {Yadav}, R.~K. 2018, \mnras, 479, 2351

\end{thebibliography}

\end{document}